# Magnetoconductivity of icosahedral and amorphous Al-Pd-Re films – a comparison

R. Haberkern [*,a], R. Rosenbaum[b], H. Bekar[b], M. Pilosof[b], A. Milner[b], A. Gerber[b], P. Häussler[a]

[a] Institut für Physik, Technische Universität Chemnitz, 09107 Chemnitz, Germany
[b] School of Physics and Astronomy, Raymond and Beverly Sackler Faculty of Exact Sciences, Tel Aviv University, Ramat Aviv, 69978, Israel

**Abstract**

The magnetoconductivity of quasicrystals is often discussed in the frame of quantum corrections, namely weak (anti-) localization and electron-electron interaction. A premise for both effects is a strong elastic scattering of conduction electrons. Amorphous and icosahedral phases are discussed as Hume-Rothery alloys with an electronically induced structural peak at the diameter of the Fermi sphere. Therefore, both should exhibit quantum corrections. The preparation of quasicrystalline films via the amorphous route offers the possibility to compare the magnetoconductivity on samples of identical composition but different structure. We report on magnetoconductivity measurements at temperatures between 0.2 K and 22 K and for magnetic fields up to 16 T. With the exception of the electronic diffusion constant, amorphous as well as icosahedral Al-Pd-Re films can be described by nearly the same set of parameters if the samples are well on the metallic side of the metal-insulator transition.

*Keywords*: Magnetoconductivity; Amorphous films; Icosahedral films; Metal-insulator transition; Hume-Rothery stabilization

## 1. Introduction

Quantum corrections to the resistivity are well known for amorphous systems in the two-dimensional (2D) [1] or three-dimensional (3D) cases [2,3]. The condition for the occurrence of quantum corrections is a strong elastic scattering which generates an extremely short mean free path in the range of only several atomic distances. However, the coherence length of the electronic wave function is much larger, since it is determined by inelastic scattering processes. In amorphous systems the large elastic scattering is brought about by scattering from the short or medium range order. This fact can be expressed in the condition

$$K_{pe} = 2k_F \quad (1)$$

with $k_F$ the Fermi vector and $K_{pe}$ the position of an electronically induced peak in the structure function $S(K)$ [4].

In the quasicrystalline state it is a-priori not clear where a large elastic scattering should come from, as the structure is well ordered. However, it has been shown for Al-Cu-Fe that the most intense diffraction peaks of the icosahedral structure (18, 29) and (20,32) are at the position of the electronically induced peak of the amorphous system and also coincide with the diameter of the Fermi sphere [5]. So, equation (1) is fulfilled also in the quasicrystalline phase, indicating a matching of the electronic wavelength at the Fermi energy $E_F$ with a characteristic length of the quasicrystalline structure. This may cause a resonant like elastic scattering [6] of the conduction electrons which leads to the occurrence of quantum corrections and to transport anomalies in the case of a moderate scattering. For a extremely strong scattering as in some samples of *i*-Al-Pd-Re this effect may result in a metal-insulator transition (MIT) [7]. Since the electronic states near $E_F$ are no longer extended but localized, the Bergmann concept of *weak* localization based on the backscattering due to a huge number of elastic scattering processes breaks down in the vicinity of the MIT.

It has been shown that the low temperature behaviour of barely metallic quasicrystals like *i*-Al-Cu-Fe and *i*-Al-Pd-Mn can be explained in terms of weak localisation (WL) and electron-electron interaction (EEI) theories [8,9,10]. In the case of (probably) insulating *i*-Al-Pd-Re Poon et al. [7] mentioned that the magnetoconductivity (MC) could not be explained in the framework WL and EEI theories, while Rodmar et al. [11,12] postulated their applicability for samples at least up to a conductivity ratio $\sigma_{295K}/\sigma_{4.2K} = 10$. Therefore, they were forced to use a strongly sample dependent spin-orbit scattering time, which to our opinion is unlikely.

Here we want to compare the conductivity as a function of temperature $T$ and magnetic field $B$ for amorphous and quasicrystalline 3D-films on the metallic side of the MIT and discuss them in the framework of WL and EEI theories.

## 2. Film Preparation and Characterization

Al-Pd-Re films were prepared by co-sputtering with two magnetron sources, one with a Re target and the other with a sectional target of Al and Pd. Quartz glass was used as substrate, held at room temperature during the deposition of the amorphous films with a thickness of 220 nm. Due to the positions of the two sources in respect to the substrate, the defined composition gradient was achieved along the substrate holder. With this technique, in one preparation process, a set of amorphous samples can be produced consisting of about 20 samples with a composition, slightly and systematically changing from one sample to the next, cutting the ternary phase diagram at, or close to the optimal composition of the quasicrystalline phase. The difference in Re-content between the insulating (in) and the metallic film (m) is 0.8 at% while the composition of the insulating film is located near $Al_{72}Pd_{20.5}Re Re_{7.5}$ as estimated by elastic recoil detection and Rutherford back-scattering.

In order to induce a transition from amorphous to quasicrystalline state, samples were annealed at 950 K in high vacuum ($p_{950K} < 5*10^{-8}$ mbar) for about 10 h. As resistance was measured during the transition, the annealing was stopped when the resistance saturated. X-ray diffraction, SEM and TEM showed [13] that the resulting films considered here are single phased icosahedral with grain sizes up to 1 µm.

## 3. Results and Discussion

Films may be classified electronically as being either insulating or (poorly) metallic. Insulating 3D films exhibit infinite resistivity or zero conductivity $\sigma$ at $T=0$. In contrast, 3D films are called poorly metallic if they display non-zero positive conductivity at $T=0$. An useful technique to identify the MIT was previously introduced [14]. The mathematical function $w(T)$:

$$w(T) = d\ln\sigma/d\ln T = (T/\sigma)d\sigma/dT , \qquad (2)$$

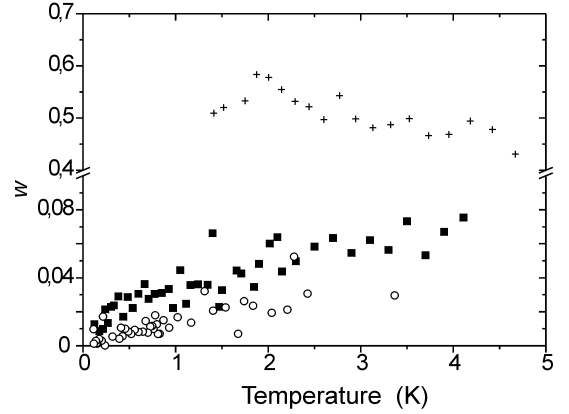

Fig.1: The $w = d\ln\sigma/d\ln T$ dependence upon temperature for the icosahedral AlPdRe films $f_{in}$ (+), $f_{bi}$ (■) and $f_m$ (○)

exhibits distinctively different temperature behaviours for insulating and metallic films.

For *strongly insulating* films exhibiting variable-range hopping, with

$$\sigma(T) = \sigma_0[\exp{-(T_0/T)^y}], \qquad (3)$$

where $\sigma_0$ is the prefactor, $T_0$ is a characteristic temperature, and $y$ is an exponent, inserting eq. (3) into eq. (2) yields:

$$w(T) = y(T_0/T)^y . \qquad (4)$$

Notice that $w(T)$ *increases* to infinity as the temperature approaches absolute zero.

In contrast, the conductivity of a *3D metallic* film at sufficiently low temperatures can be described by a power law:

$$\sigma(T) = \sigma(0) + CT^z , \qquad (5)$$

where $\sigma(0)$ is the positive conductivity at $T=0$ and $C$ is the prefactor of the power law term. Eq. (5) might approximate the conductivity contribution of critical states or from the 3D EEI and WL theories. Inserting eq. (5) into eq. (2) yields

$$w(T) = zCT^z/[\sigma(0) + CT^z] = zCT^z/\sigma(T) \qquad (6)$$

If the film is *metallic* and exhibits a finite positive conductivity $\sigma(T=0)$, then $w(T)$ should extrapolate to *zero* for $T \rightarrow 0$ K.

Fig.1 shows the $w(T)$ behaviour for 3 different icosahedral films ($f_{in}$, $f_m$, $f_{bi}$). Film $f_m$ exhibits $w$-values which tend to zero for $T \rightarrow 0$ K and is clearly metallic. In contrast, the $w$ behaviour exhibited by Film $f_{in}$ suggests that the w-values extrapolate to a *finite value* or even increase down to smaller temperatures. Hence, this film appears to be insulating. Film $f_{bi}$ seems to be in the vicinity of a MIT as its $w$-values tend to a very small but probably finite value for $T \rightarrow 0$.

Low-$T$ MC data are plotted in fig. 2 to 5 for the 3 different icosahedral and an amorphous film $f_{am}$. From a qualitative point of view all *metallic* samples regardless if amorphous or icosahedral exhibit exclusively negative

MC for all measured temperatures. This is as expected for films containing high Z elements like Pd and especially Re due to strong spin-orbit scattering. The positive MC at $B<3$ Tesla and $T=4.2$ K for film $f_{in}$ (fig. 2) is typical for *i*-Al-Pd-Re samples on the insulating side of the MIT and can not be interpreted using the theory of WL.

As summarized in fig. 6 for 20 different films the absolute change of the conductivity $\Delta\sigma$ by a magnetic field $B=8$ Tesla at $T=1.3$K is very similar in size and depends only slightly from the conductivity which changes over 3 orders of magnitude. Especially, no significant change occurs at the transition from amorphous to icosahedral Al-Pd-Re films. Thereby, weak (anti-) localization contributes about 80% to the negative MC magnitude. In the high field limit it is proportional to $D^{-1/2}$ ($D$: electronic diffusion const.) and a function of the spin-orbit $\tau_{so}$ and inelastic scattering times $\tau_{ie}(T)$. As both scattering times probably do not show a considerable change with structure, fig. 6 suggests that also the diffusion constant is nearly constant crossing the transition from amorphous to icosahedral structure. This may be astonishing as the conductivity changes by a factor of $10^2$ but is in accordance with measurements of the Hall mobility [15] which also shows a comparably small change.

For a quantitative analysis, the MC data for the metallic icosahedral film $f_m$ were fitted using the contributions arising from EEI in the particle-hole [16] and the 3D WL including spin-orbit scattering and Zeeman splitting [17]. According to the lack of any other better formalism, we use the WL theory close to the MIT for film $f_m$; the WL theory generally applies to highly metallic films.

A magnitude for the spin-orbit scattering time [18] was estimated from the expression $\tau_{SO} \approx \tau_0 (137/Z)^4$ to be $\tau_{so} \approx 10^{-13}$ s, where $Z$ is the atomic number and $\tau_0 \approx 10^{-15}$s is the elastic scattering time. The resulting fitting parameters are listed in Table 1 and are quite similar to the values reported in [11,12] extracted from the *metallic* bulky i-samples. The inelastic scattering times yielded from the MC fit at different $T$ could be fitted to a simple power law $\tau_{ie}(T)=\tau_{ie0} T^p$ where $p\approx-1$. The MC fits describe the 2 metallic of the icosahedral films quite well while the MC of the amorphous film at $T=2.15$K is strongly influenced from superconducting fluctuations which are a precursor to the superconducting transition at $T=0.55$K.

Table 1: Fitting parameters to the MC data, $\tau_{so}$ and $\tau_{ie}$ denote the spin-orbit and inelastic scattering times, respectively, $F$ the electron screening parameter, $D$ the electronic diffusion constant and $p$ the exponent of the power law for inelastic scattering times

| film | $\tau_{so}$ [s] | $\tau_{ie}$ (4.2 K) [s] | $p$ | $F$ | $D$ [cm$^2$/s] |
|---|---|---|---|---|---|
| $f_m$ | 1.5e-13 | 3.6e-12 | -1.03 | 0.2 | 0,75 |
| $f_{bi}$ | 1.5e-13 | 1,8e-11 | -1.03 | 0.2 | 0,15 |
| $f_{am}$ | 1.5e-13 | 5.5e-11 | -2 | 0.2 | 1,5 |

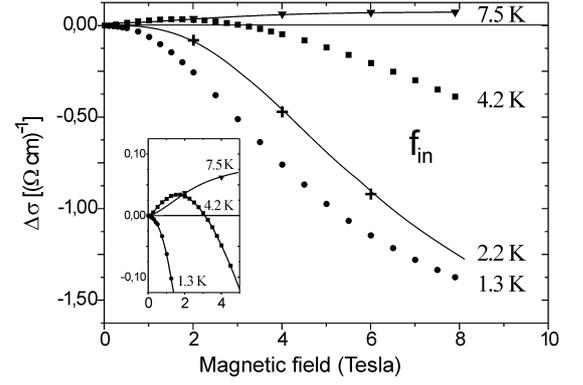

Fig. 2: Magnetoconductivity of the probably insulating film $f_{in}$ as a function of magnetic field

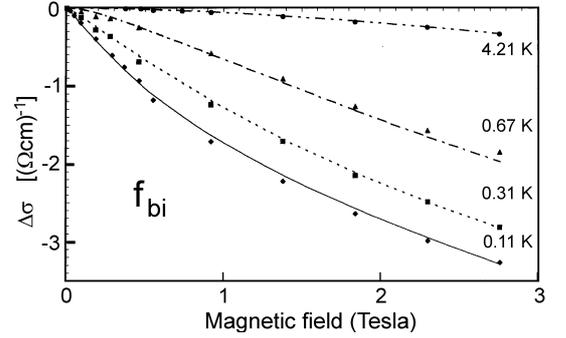

Fig.3: Magnetoconductivity of the barely insulating film $f_{bi}$ as a function of magnetic field; lines show a fit according to WL and EEI contributions

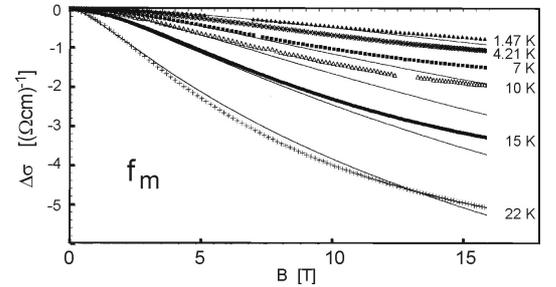

Fig.4: Magnetoconductivity of the "metallic" film $f_m$ as a function of magnetic field; lines show a fit according to WL and EEI contributions

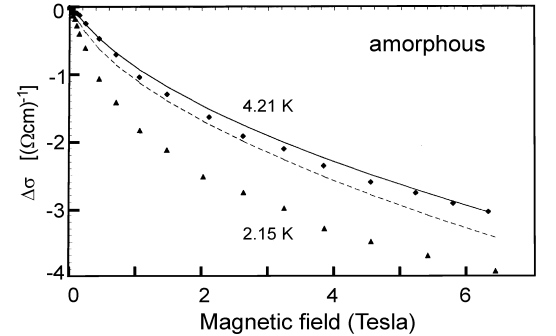

Fig.5: Magnetoconductivity of an amorphous film $f_{am}$ as a function of magnetic field; lines show a fit according to WL and EEI contributions

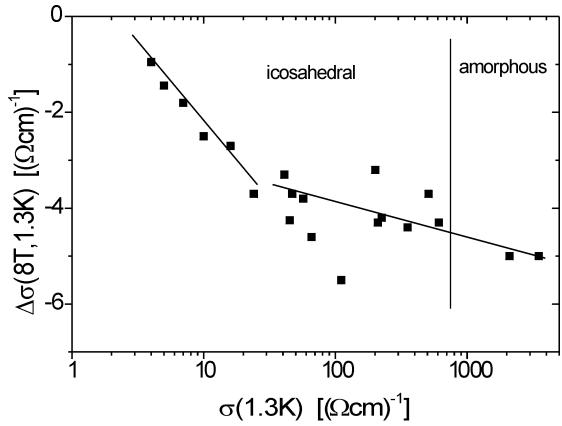

Fig.6: Magnetoconductivity at *B*=8Tesla and *T*=1.3K as a function of the conductivity at *T*=1.3K for icosahedral and amorphous Al-Pd-Re films

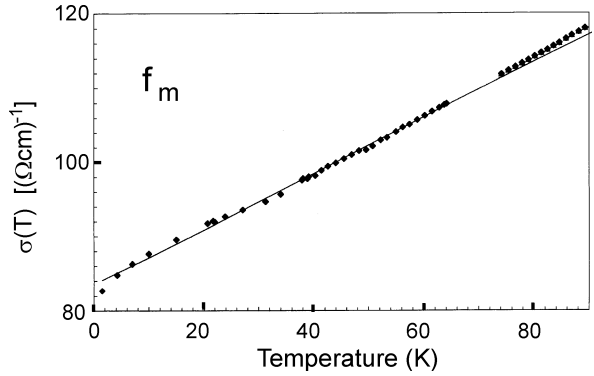

Fig.7: Zero field conductivity of the "metallic" film $f_m$; the line shows a fit according to WL and EEI contributions as obtained from the magnetoconductivity data

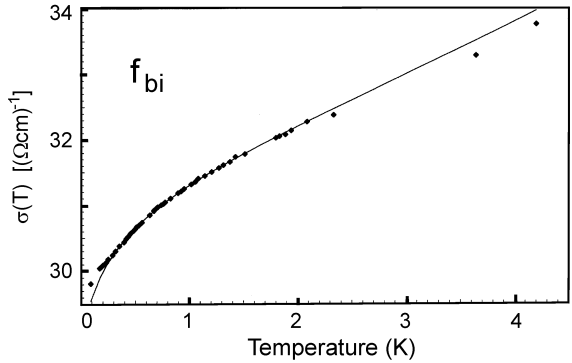

Fig.8: Zero field conductivity of the barely insulating film $f_{bi}$; the line shows a fit according to WL and EEI contributions as obtained from the magnetoconductivity data

The *zero field* conductivity data of *i*-films $f_m$ and $f_{bi}$ are compared to data computed by using the parameters which were determined from the MC fits. The single parameter fit of $\sigma(T)=\sigma(T=0)+\Delta\sigma_{WL}+\Delta\sigma_{EEI}$ is impressively good for film $f_m$ between 4 K to 60 K with $\sigma(T=0) = 18.7$ $(\Omega cm)^{-1}$ as shown in fig. 7. In this case, the EEI expression contributes 70 % to the total conductivity change with temperature, while the WL expression makes a smaller 30 % contribution. Interestingly, the WL expression exhibits anti-localization below 6 K.

For the barely insulating film $f_{bi}$ (fig.8), initial inspection indicates agreement between data and theory. But the fit is unphysical since a negative conductivity $\sigma(T=0)=-85(\Omega cm)^{-1}$ had to be used. Thus the WL contribution greatly overestimates the zero field conductivity; and the WL theory breaks down already in the vicinity of the MIT. The above illustration marks the importance of fitting both MC and zero field conductivity data only to metallic film and checking for consistency between all the fits.

## 4. Conclusions

Magnetoconductivity as well as zero field conductivity data of icosahedral as well as amorphous Al-Pd-Re films can be described by WL and EEI theories if the films are well on the metallic side of the metal-insulator transition. Besides the electronic diffusion constant, amorphous and icosahedral films of very different conductivities can be described by nearly the same set of parameters.

## Acknowledgements

We greatly thank the Tel Aviv University Internal Research Fund and DFG for its financial support.

## References

[1]  G. Bergmann, Phys. Reports 107 (1984) 30.
[2]  H. Fukuyama, K. Hoshino, J. Phys. Soc. Jpn. 50 (1981) 213.
[3]  B.L. Al'tshuler, A.G. Aronov, JETP Lett. 30 (1979) 482.
[4]  P. Häussler, Physic. Rep. 222 (2) (1992) 65.
[5]  C. Roth, G. Schwalbe, R. Knöfler, F. Zavaliche, O. Madel, R. Haberkern, P. Häussler, J. Non-Cryst. Solids 250-252 (1999) 869.
[6]  P. Häussler, H. Nowak, R. Haberkern, this volume.
[7]  S.J. Poon, F.S. Pierce, Q. Guo, Phys. Rev. B 51 (1995) 2777.
[8]  A. Sahnoune, J.O. Ström-Olsen, A. Zaluska, Phys. Rev.B 46 (1992) 10629.
[9]  R. Haberkern, G. Fritsch and J. Schilling, Z. Phys. B 92 (1993) 383.
[10]  P. Lindqvist, P. Lanco, C. Berger, A.G.M. Jansen, F. Cyrot-Lackmann, Phys. Rev. B 51 (1995 ) 4796 .
[11]  M. Rodmar, Oberschmidt, M. Ahlgren, C. Gignoux, C. Berger Ö. Rapp, J. Non-Cryst. Solids 250-252 (1999) 883.
[12]  M. Ahlgren, M. Rodmar, C. Gignoux, C. Berger and Ö. Rapp, Mat. Sci. Eng. A 226-228 (1997) 981.
[13]  R. Haberkern, C.Roth, R.Knöfler, S.Schulze, P. Häussler, MRS Symposium Proceedings Vol. 553 (1999) p. 13.
[14]  A. Möbius, Phys. Rev. B 40 (1989) 4194.
[15]  R. Haberkern, K. Khedhri, C. Madel, P. Häussler, this volume.
[16]  P.A. Lee, T.V. Ramakrishnan, Rev. Mod. Phys. 57 (1985) 308.
[17]  D.V. Baxter, R. Richter, M.L. Trudeau, R.W. Cochrane, J.O. Strom-Olsen, J. Physique 50 (l989) 1673
[18]  A.A. Abrikosov and L.P. Gor'kov, Zh. Eskp. Teor. Fiz. 42 (1962) 1088 [Sov. Phys. - JETP 15 (1962) 752.]